\documentclass[12pt]{imsart}

\usepackage{enumitem}
\usepackage[utf8]{inputenc}
\usepackage[english]{babel}
\usepackage[margin=1in]{geometry}
\usepackage[colorlinks,citecolor=blue,urlcolor=blue]{hyperref}

\usepackage{bm}
\usepackage{graphicx}
\usepackage{amssymb,amsmath,amsthm,mathrsfs}
\usepackage{epstopdf}
\usepackage{algorithmic}
\usepackage{xcolor}
\usepackage{subfigure}
\usepackage{multirow}
\usepackage{titlesec,textcase}
\usepackage{longtable}
\usepackage{rotating}
\usepackage{booktabs}
\usepackage{siunitx} 
\usepackage{float}

\newtheorem{Theorem}{Theorem}[section]

\newtheorem{Definition}[Theorem]{Definition}

\theoremstyle{definition}

\RequirePackage[normalem]{ulem} 

\definecolor{rp}{RGB}{83,54,106}

\def\boxit#1{\vbox{\hrule\hbox{\vrule\kern6pt\vbox{\kern6pt#1\kern6pt}\kern6pt\vrule}\hrule}}

\begin{document}
\begin{frontmatter}
\title{Asymptotic distribution of the global clustering coefficient in a random annulus graph}

\runtitle{Clustering coefficient of RAG }
\runauthor{ M. Yuan }
 
\begin{aug}
\author[B]{\fnms{Mingao} \snm{Yuan}\thanks{Mingao Yuan is grateful for the generous support provided by the Startup Fund of The University of Texas at El Paso.}\ead[label=e2]{myuan2@utep.edu}}

\address[B]{Department of Mathematical Sciences,
The University of Texas at El Paso, El Paso, TX, USA\\
\printead{e2}}

\end{aug}
 
\begin{abstract}
The global clustering coefficient is an effective measure for analyzing and comparing the structures of complex networks. The random annulus graph is a modified version of the well-known  Erd\H{o}s-R\'{e}nyi random graph. It has been recently proposed in modeling network communities. This paper investigates the asymptotic distribution of the global clustering coefficient in a random annulus graph. It is demonstrated that the standardized global clustering coefficient converges in law to the standard normal distribution. The result is established using the asymptotic theory of degenerate U-statistics with a sample-size dependent kernel. As far as we know, this method is different from established approaches for deriving asymptotic distributions of network statistics. Moreover, we get the explicit expression of the limit of the global clustering coefficient.
\end{abstract}



\end{frontmatter}

\section{Introduction}

A network or graph is a mathematical model that captures the pairwise interactions between a group of individuals. While traditional datasets capture individual characteristics, network data records the relationships between pairs of individuals. In a network of $n$ individuals, relationships are represented by an $n\times n$ adjacency matrix. The entry at $(i,j)$  models the interaction between individual $i$ and individual $j$. Network data analysis has diverse applications across various scientific disciplines, encompassing both the social sciences and the biological sciences \cite{N03,BM08,SBL13,CBL16,CGL16}.

A fundamental objective in network data analysis is to characterize the structural topology of a network. Various summary metrics have been formulated to quantify the network topology. These metrics facilitate the comparative analysis of networks and their categorization according to salient structural features. The global clustering coefficient is among the most frequently applied graph statistics \cite{WS98,N09,FHW23}. It is defined as the number of closed triplets over the total number of triplets. The global clustering coefficient measures the extent to which nodes in a graph tend to cluster together.

The random annulus graph is a generalization of the random geometric graph  \cite{GPAS18,GMPS23,DC23,LM24}. It incorporates the Goldilocks principle of real world networks, wherein interactions are strongest within an intermediate range of proximity--neither too close nor too far \cite{LM24}. The random annulus graph model also arises in the context of community detection for random geometric graphs \cite{GPAS18,GMPS23}. In the random annulus graph, nodes are randomly and uniformly placed in  the interval [0,1]. Two nodes are connected by an edge if their distance is less than $r_1$ and greater than $r_2$. When $r_2=0$, the random annulus graph is reduced to the random geometric graph \cite{DC23}.  The connectivity properties of random annulus graphs are investigated in \cite{GPAS18,GMPS23}.  \cite{LM24} shows the deterministic annulus graphs are uniquely determined by the ratio $r_1/r_2$ when $r_2=o(r_1)$.

In network analysis, a significant area of research is dedicated to investigating the asymptotic properties of summary statistics \cite{BM06,CHHS21,Y23,Y23b,Y23c}. For example, \cite{BM06} derived the average number of cliques in random scale-free networks, while \cite{CHHS21} derived the asymptotic distribution of the empirical distribution of  a random graph's adjacency and Laplacian matrices. Additionally, \cite{Y23b} established the limiting distributions of  a class of topological indices.

In this paper, we study the asymptotic distribution of the global clustering coefficient in  the random annulus graph.  We prove that the suitably scaled and centered global clustering coefficient converges in distribution to the standard normal distribution. The proof is based on an innovative application of the theory of degenerate U-statistics with sample-size-dependent kernel. Moreover, our analysis also shows that the global clustering coefficient is asymptotically equal to $\frac{3}{4}\frac{(r_1-2r_2)^2}{(r_1-r_2)^2}$. This result contrasts with the corresponding random geometric graph, for which the limit is  $\frac{3}{4}$. In this sense, the random annulus graphs may offer a more accurate model for real-world networks than the corresponding random geometric graphs, due to their ability to produce networks with a global clustering coefficient between 0 and $\frac{3}{4}$.

The remainder  of the paper is organized as follows. Section 2 presents the definition of the global clustering coefficient, the random annulus graph model, and the main result. Section 3 presents the proof.

Notations: Let $c_1,c_2$ be two positive constants. For two positive sequence $a_n$, $b_n$, denote $a_n=\Theta( b_n)$ if $c_1\leq \frac{a_n}{b_n}\leq c_2$; denote $a_n=O(b_n)$ if $\frac{a_n}{b_n}\leq c_2$; $a_n=o(b_n)$ if $\lim_{n\rightarrow\infty}\frac{a_n}{b_n}=0$. Let $X_n$ be a sequence of random variables. We use $X_n\Rightarrow F$ to denote $X_n$ converges in distribution to a probability distribution $F$. $X_n=O_P(a_n)$ means $\frac{X_n}{a_n}$ is bounded in probability.  $X_n=o_P(a_n)$ means $\frac{X_n}{a_n}$ converges to zero in probability. $N(0,1)$ stands for the standard normal distribution. The notation $\sum_{i\neq j\neq k}$ represents summation over indices $i,j,k$ with $i\neq j, j\neq k, i\neq k$.

\section{ Asymptotic distribution of the global
clustering coefficient}\label{main}

A graph or network consists of a nodes (vertices) set and an edge set. Let $\mathcal{V}=\{1,2,\dots,n\}$ for positive integer $n$. 
An \textit{undirected} graph on $\mathcal{V}$ is a pair $\mathcal{G}=(\mathcal{V},\mathcal{E})$, where $\mathcal{E}$ is a set of edges. An edge connects two nodes and is a subset of $\mathcal{V}$ with cardinality two.  For $i<j$, denote $A_{ij}=1$ if $\{i,j\}$ is an edge and $A_{ij}=0$ otherwise. Suppose $A_{ij}=A_{ji}$. Then the symmetric matrix $A=[A_{ij}]$ is called adjacency matrix of graph $\mathcal{G}$.  A triangle is a set of three vertices where each pair is connected by an edge. A 2-path is a sequence of two edges that joins three distinct vertices.

The global clustering coefficient $\mathcal{C}_n$ of a graph is defined as \cite{WS98,N09}
\[\mathcal{C}_n=\frac{\sum_{i\neq j\neq k}A_{ij}A_{jk}A_{ki}}{\sum_{i\neq j\neq k}A_{ij}A_{jk}}.\]
The fraction is the number of triangles multiplied by three, divided by the number of 2-paths multiplied by  two. It measures the degree to which nodes in a graph tend to cluster together.

A graph is said to be random if  the presence or absence of edges between nodes is determined probabilistically. In the well-known Erd\H{o}s-R\'{e}nyi 
random graph, each edge has the same fixed probability of being present or absent, independently of the other edges. In a random geometric graph,  nodes are randomly placed in some metric space and  two nodes are connected by an edge if and only if their distance is smaller than a certain  radius  \cite{DC23}. In this paper, we are interested in the random annulus graphs defined below \cite{GPAS18,GMPS23}.

\begin{Definition}\label{def1}
Let $r_1,r_2\in[0,0.5]$ be real numbers such that $r_1>r_2$. Given i.i.d. uniform random variables $X_1,X_2,\dots,X_n$ on the interval $[0,1]$, the Random  Annulus Graph (RAG) $\mathcal{G}_{n}(r_1,r_2)$ is defined as
\[A_{ij}=I[r_2<d(X_i,X_j)< r_1],\]
 where $A_{ii}=0$, and 
\begin{eqnarray}\label{distance}
 d(X_i,X_j)=\min\{|X_{i}-X_{j}|,1-|X_{i}-X_{j}|\}.
 \end{eqnarray}
\end{Definition}

In $\mathcal{G}_{n}(r_1,r_2)$, the nodes are independently and uniformly placed in the interval $[0,1]$, and an edge connects two nodes if their distance is less than $r_1$ and greater than $r_2$. The distance $ d(X_i,X_j)$ is the distance on a circle with circumference 1 \cite{BB24}. When $r_2=0$, the random  annulus graph is reduced to the random geometric graph.  Random annulus graphs appear naturally in modeling communities in networks \cite{GPAS18,GMPS23}. 

A major research theme is the asymptotic properties of network statistics \cite{CHHS21,Y23b}.  This work analyzes the limiting distribution of the global clustering coefficient of the random annulus graphs.

\begin{Theorem}\label{thm1} Suppose $A\sim \mathcal{G}_{n}(r_1,r_2)$ with $2r_2<r_1=O(r_2)$, $r_1=o(1)$ and $nr_1=\omega(1)$. Then
\begin{equation}\label{clusdist}
 \frac{2\sqrt{2}(r_1-r_2)^2n}{3\sigma_{n2}}\left(\mathcal{C}_n-\frac{3}{4}\frac{(r_1-2r_2)^2}{(r_1-r_2)^2}\right)\Rightarrow N(0,1),
\end{equation}
where
\[\sigma_{n2}^2=\mathbb{E}[h(X_1,X_2,X_3)h(X_1,X_2,X_4)]=\Theta(r_1^3),\]
\begin{eqnarray*}
h(X_1,X_2,X_3)=A_{12}A_{13}A_{23}-\frac{(r_1-2r_2)^2}{4(r_1-r_2)^2}(A_{12}A_{13}+A_{21}A_{23}+A_{31}A_{32}).
\end{eqnarray*}
\end{Theorem}

\medskip

In Theorem \ref{thm1}, the condition $r_1=o(1)$ means the network is sparse. This is a reasonable assumption, as most real-world networks exhibit sparsity \cite{A17}. When $2r_2\leq r_1$, the probability of forming a triangle is zero. When $r_1=\omega(r_2)$, the  probability of forming an edge in RAG is almost the same as the random geometric graph. Therefore, we assume $2r_2<r_1=O(r_2)$. The condition $nr_1=\omega(1)$ requires that the network cannot be excessively sparse. This is a common and minor constraint \cite{GPAS18,GMPS23} .

The proof of Theorem \ref{thm1} is non-trivial because the lack of independence among the edges of $\mathcal{G}_{n}(r_1,r_2)$ requires a more intricate approach. Our approach involves expressing $\mathcal{C}_n$ as a degenerate U-statistic whose kernel function varies with the network size $n$. Subsequently, the asymptotic theory of degenerate U-statistics with sample-size-dependent kernels is employed to establish its limiting distribution.

According to Theorem \ref{thm1}, the standardized global clustering coefficient converges in distribution to the standard normal distribution as the network size $n$ goes to infinity. Moreover, equation (\ref{clusdist}) implies that 
\begin{eqnarray}
\mathcal{C}_n=\frac{3}{4}\frac{(r_1-2r_2)^2}{(r_1-r_2)^2}+O_P\left(\frac{1}{n}\right).
\end{eqnarray}
That is, the global clustering coefficient is asymptotically equal to $\frac{3}{4}\frac{(r_1-2r_2)^2}{(r_1-r_2)^2}$. Let $r_1=\lambda r_2$ with a constant $\lambda>2$. Then 
\begin{eqnarray}\label{gccal}
0<\frac{3}{4}\frac{(r_1-2r_2)^2}{(r_1-r_2)^2}=\frac{3}{4}\frac{(\lambda-2)^2}{(\lambda-1)^2}<\frac{3}{4}.
\end{eqnarray}
Asymptotically, $\mathcal{C}_n$ can take any value between 0 and $\frac{3}{4}$. Note that $\mathcal{G}_{n}(r_1,0)$ is the random geometric graph. The global clustering coefficient of $\mathcal{G}_{n}(r_1,0)$ is asymptotically equal to $\frac{3}{4}$. Therefore, $\mathcal{G}_{n}(r_1,r_2)$ has properties that are markedly different from the corresponding random geometric graph $\mathcal{G}_{n}(r_1,0)$.


\begin{thebibliography}{9}


\bibitem{A17} Abbe, E. (2017).
Community detection and stochastic block models: recent developments. 
\textit{Journal of Machine Learning Research}, \textbf{18}, 1-86.



\bibitem{BM06}
Bianconi, G. and Marsili, M. (2006).
Number of cliques in random scale-free network ensembles,\textit{Physica D: Nonlinear Phenomena}, 224,:1-6.











\bibitem{CHHS21}
Chakrabarty, A., Hazra, S. R., Hollander, F. D. and Sfragara, M.(2021).
Spectra of adjacency and Laplacian matrices of inhomogeneous  Erd\"{o}s-R\'{e}nyi random graphs,
\textit{Random matrices: Theory and applications}, 10(1),215009.

















\bibitem{BBA01}
Becq, G. J. C., Barbier, E. L., Achard, S. (2020). Brain networks of rats under anesthesia using resting-state fMRI: comparison with dead rats, random noise and generative models of networks. Journal of Neural Engineering, 17(4), 045012.

\bibitem{BB24}
Bangachev,K. and Bresler, G.(2024).
Detection of $L_{\infty}$ geometry in random geometric graphs:suboptimality of triangles and cluster expansion. \textit{Proceedings of Machine Learning Research}, 247:1–71.







 

    
 


\bibitem{CGL16}
 Chiasserini, C.F.,  Garetto, M. and  Leonardi, E. (2016). Social network de-anonymization under
scale-free user relations. \textit{IEEE/ACM Transactions on Networking} 24 (6):3756–3769.

\bibitem{CBL16}
Charitou, T., Bryan, K. and Lynn, D.(2016).
Using biological networks to integrate, visualize and analyze genomics data,\textit{Genetics Selection Evolution}, 48,27.




\bibitem{DC23}
Duchemin, Q., De Castro, Y. (2023). Random geometric graph: some recent developments and perspectives. \textit{High Dimensional Probability IX. Progress in Probability} Birkhäuser, Cham.  





\bibitem{FL96}
Fan, Y. and Li, Q. (1996). Consistent model specification tests: omitted variables and semiparametric functional
forms.\textit{Econometrica},  64, 4,865-890.


 

\bibitem{FHW23}
Fatima, U., Hina, S. and Wasif, M.(2023). A novel global clustering coefficient-dependent degree centrality (GCCDC)
metric for large network analysis using real-world datasets, 
\textit{Journal of Computational Science }, 70, 102008.


\bibitem{GPAS18}
Galhotra, S.,   Mazumdar, A., Pal, S., Saha, B.(2023).
The geometric block model and applications, 2018 56th Annual Allerton Conference on Communication, Control, and Computing (Allerton), Monticello, IL, USA.


\bibitem{GMPS23}
Galhotra, S.,   Mazumdar, A., Pal, S., Saha, B.(2023).
Community recovery in the geometric block model, \textit{Journal of Machine Learning Research} 24: 1-53.


\bibitem{LM24}
Lichev, L. and Mihaylov, T.(2024). Annulus graphs in $R^d$. \textit{Discrete \& Computational Geometry}, 72, 379-401.



    \bibitem {N03} Newman, M E J (2003),
    The structure and function of complex networks. \textit{SIAM Review} 45, 167-256 .

\bibitem{N09}
Newman, M.E.J. (2009). Random Graphs with Clustering, \textit{Physical Review Letters}, 103, 058701.



\bibitem{BM08}
 O'Malley, A. J. and Marsden,P. V. (2008). The analysis of social networks, \textit{Health Serv Outcomes Res Methodol}, {\bf 8}, 222-269.









\bibitem{SBL13}
Simpson, S., Bowman F. and Laurienti, P.(2013). Analyzing complex functional brain networks: Fusing statistics and network science to understand the brain,\textit{Statistics Surveys}, 7: 1-36.



\bibitem{WS98}
Watts,D. and Strogatz, S. (1998).
Collective dynamics of ‘small-world’ networks, \textit{Nature}, 393, 440-442. 


\bibitem{Y23}
M. Yuan,  {\it On the Randić index and its variants of network dat},  TEST, {\bf 33} (2024), 155–179.



\bibitem{Y23b}
M. Yuan,  {\it Asymptotic distribution of degree-based topological indices},  MATCH Commun. Math. Comput. Chem., {\bf 91:1} (2023), 135-196.


\bibitem{Y23c}
M. Yuan,  {\it On the Renyi index of random graphs}, Statistical Papers, {\bf 65} (2024),  1773–1803.

    \bibitem{Y25}
    Yuan, M. (2025). Hypothesis testing for the dimension of
random geometric graph, preprint, ResearchGate, DOI: 10.13140/RG.2.2.31959.53920
 
    

 

    





\end{thebibliography}
\end{document}